\documentstyle[preprint,aps]{revtex}
\draft
\begin{document}
\date{\today}
\title{
Composite Fermions and the Energy Gap in the Fractional Quantum Hall
Effect}
\author{N.~E.~Bonesteel}
\address{
National High Magnetic Field Laboratory, Florida State University,
Tallahassee, FL 32306-4005}
\maketitle
\begin{abstract}
\noindent
The energy gaps for the fractional quantum Hall effect at filling
fractions 1/3, 1/5, and 1/7 have been calculated by variational Monte
Carlo using Jain's composite fermion wave functions before and after
projection onto the lowest Landau level. Before projection there is a
contribution to the energy gaps from the first excited Landau level.
After projection this contribution vanishes, the quasielectron charge
becomes more localized, and the Coulomb energy contribution increases.
The projected gaps agree well with previous calculations, lending
support to the composite fermion theory.
\end{abstract}

In both the integer and fractional quantum Hall effect (IQHE and FQHE) a
two dimensional electron gas in a strong perpendicular magnetic field
$B$ exhibits a precisely quantized Hall resistance and an exponentially
activated longitudinal resistance $R_{xx}\propto\exp-\Delta/2KT$
indicating the appearance of an energy gap $\Delta$ in the system
\cite{iqhe,fqhe,qhe}.  In the IQHE $\Delta$ is equal to the energy
required to promote an electron from the highest filled Landau level to
the lowest empty Landau level, which in a weakly disordered system with
fully polarized spins is approximately the cyclotron energy:
$\Delta\simeq\hbar\omega_c\equiv\hbar eB/m^*c$ where $m^*$ is the band
mass of the electron.  This is to be contrasted with the FQHE where an
energy gap forms in a partially filled Landau level due to correlation
effects \cite{laughlin} and $\Delta\propto e^2/\epsilon l_0$ where
$\epsilon$ is the dielectric constant and $l_0 = \sqrt{\hbar c/eB}$ is
the magnetic length.

Jain has introduced the concept of composite fermions, electrons bound
to an even number of flux quanta, in order to treat the IQHE and FQHE on
the same footing \cite{jain}.  According to Jain's theory composite
fermions experience an effective magnetic field $B_{\rm eff} = B -
2k\phi_0n$ where $\phi_0 = hc/e$ is the flux quantum, $2k$ is the number
of bound flux quanta, and $n$ is the electron density.  When composite
fermions fill $p$ {\it pseudo}-Landau levels, where a pseudo-Landau
level is a Landau level corresponding to the effective magnetic field
$B_{\rm eff}$, the physical Landau level filling fraction is
$\nu=p/(2kp+1)$ and the system exhibits a FQHE with $\Delta$ equal to
the energy required to promote a composite fermion from the $p$th to the
$(p+1)$th pseudo-Landau level.  The energy gap for the FQHE can
therefore be thought of as an effective cyclotron energy for composite
fermions.  Du {\it et al.} \cite{du} have measured the energy gaps for
the sequence $\nu=p/(2p+1)$ and found that $\Delta(B) \propto |B_{\rm
eff}| = |B - B_{1/2}|$ where $B_{1/2} = 2\phi_on$, consistent with
Jain's theory
\cite{comment1}.  In addition, a number of experiments exploring the
regime near $\nu\sim 1/2$ where $B_{\rm eff} \sim 0$ have provided
evidence for the existence of a composite fermion `metal' in this limit
\cite{kang,halperin}.  These experiments have led to a renewed interest
in composite fermions.

In this paper I present calculations of the energy gaps for the
$\nu=1/3$, 1/5, and 1/7 FQHE using Jain's composite fermion wave
functions \cite{jain}.  Before proceeding it is useful to contrast the
present work with that of Morf and Halperin \cite{morf} who calculated
$\Delta$ for $\nu=1/3$ by first calculating the excitation energies for
an isolated quasielectron $E_{\rm qe}$ and quasihole $E_{\rm qh}$ using
Laughlin's trial wave functions in the disk geometry.  Morf and Halperin
found that $\Delta = E_{\rm qe}+E_{\rm qh} \simeq 0.1 e^2/\epsilon l_0$,
a result which agrees well with experiment once the effects of Landau
level mixing \cite{yoshioka} and the finite thickness of the wave
function \cite{zhang} are properly taken into account \cite{willett}.
Note that although the quasielectron and quasihole form a bound state
(the magnetoroton) with lower energy than $\Delta$ this state does not
carry current and so does not contribute to $R_{xx}$.  In Jain's theory
the ground state for $\nu=1/(2k+1)$ is obtained by completely filling
the lowest pseudo-Landau level with composite fermions. A quasihole is
created by removing a composite fermion from this state, while a
quasielectron is created by adding a composite fermion to the first
excited pseudo-Landau level.  Jain has shown that the composite fermion
wave functions for the ground state and quasihole state are identical to
Laughlin's \cite{jain}.  Thus the essential difference between the
present calculation and that of Morf and Halperin is that Jain's rather
than Laughlin's quasielectron wave function is used.

Because Jain's quasielectron wave function has a nonzero overlap with
the first excited Landau level it is necessary to project onto the
lowest Landau level to study the $B\rightarrow\infty$ limit.  Previously
this has only been done on small systems (up to 10 electrons for
$\nu=1/3$) where it is possible to perform this projection exactly
\cite{kasner}.  Here it is shown how projected expectation values for
states with a single quasielectron can be expressed in terms of
unprojected expectation values which can then be evaluated by
variational Monte Carlo.  Using this projection technique it is possible
to study larger systems than have previously been studied and perform a
reliable extrapolation to the thermodynamic limit.  The main result of
this paper is that for $\nu=1/(2k+1)$ projection not only trivially
ensures that the contribution to the energy gap from the first excited
Landau level vanishes, but also has the effect of further localizing the
charge of the quasielectron, thus transforming the kinetic energy
contribution to the unprojected energy gap ($\propto\hbar\omega_c)$ into
an increased Coulomb contribution ($\propto e^2/\epsilon l_0$).  For
$\nu=1/3$ and 1/5 the projected energy gaps agree well with previous
calculations, establishing the variational validity of Jain's
quasielectron wave function.  This is a nontrivial result because even
after projection Jain's quasielectron is not equivalent to Laughlin's
\cite{jain}.

Adopting the spherical geometry introduced by Haldane \cite{haldane} the
Hamiltonian for spin polarized electrons on a sphere with a magnetic
monopole at the center is $H =T+V$ where
\begin{equation}
T = \frac{\omega_c}{2 \hbar S} \sum_{i=1}^N
\left({\bf r}_i\times\left(i\hbar{\bf\nabla}_i-\frac{e}{c}{\bf
a}_i\right)\right)^2 - \frac{N}{2}\hbar\omega_c
\end{equation}
and
\begin{equation}
V = \sum_{i<j}\frac{e^2}{\epsilon r_{ij}}.
\end{equation}
Here $2S$ is the number of flux quanta passing through the surface of
the sphere, $N$ is the total number of electrons, ${\bf a}_i = S {\bf
e}_\phi \cot\theta_i /R$ is the vector potential corresponding to the
magnetic field $B = 2S\phi_0/4\pi R^2$, and $R$ is the radius of the
sphere.  The zero point kinetic energy has been subtracted from $T$ and
the distance between electrons in $V$ is taken to be the chord distance.
The one-body eigenstates of $T$ in the lowest and first excited Landau
levels with eigenvalues $E_0=0$ and $E_1=(1+1/S)\hbar\omega_c$ have
angular momentum $l=S$ and $l=S+1$ and are given by
\begin{equation}
\Upsilon_{S,m} \propto u^{S-m}v^{S+m}
\end{equation}
for $m= -S,...,S$ and
\begin{equation}
\Upsilon_{S+1,m} \propto ((2S+2){\overline v}v+m+S+1)u^{S-m}v^{S+m}
\end{equation}
for $m=-S-1,...,S+1$ \cite{haldane}.  Here $m$ labels the $z$ component
of the angular momentum, and $u = \cos\theta/2 \exp -i\phi/2$ and $v =
\sin\theta/2\exp i\phi/2$ are the spinor coordinates of the electron.

Jain has proposed a simple relationship between the physical electron
wave function $\psi$ and the corresponding `mean field' composite
fermion wave function $\Phi_{\rm CF}$.  On the sphere this relationship
is
\begin{equation}
\psi = \prod_{i<j}(u_iv_j -v_i u_j)^{2k} \Phi_{\rm CF}\label{jain}
\end{equation}
where the Jastrow factor in (\ref{jain}) corresponds intuitively to
attaching $2k$ flux quanta to the electrons and can be derived
microscopically using the Chern-Simons theory of composite fermions
\cite{lopez}.  

Using (\ref{jain}) it is straightforward to construct the ground state
and excited state wave functions needed to calculate $\Delta$. When
$\nu=1/(2k+1)$ the number of flux quanta passing through the surface of
the sphere is related to the total number of electrons by $2S =
(2k+1)(N-1)$ \cite{haldane}.  For this field strength the effective
field seen by composite fermions with $2k$ flux quanta attached
corresponds to $2S^* = N-1$ \cite{jain}.  When the lowest pseudo-Landau
level is completely full the mean field composite fermion wave function
is
\begin{eqnarray}
\Phi_{CF} &=& \left|
\begin{array}{ccccc}
u_1^{N-1} & u_1^{N-2}v_1 & ... & u_1 v_1^{N-2} & v_1^{N-1} \\
\vdots & \vdots  &  &\vdots &\vdots  \\
u_N^{N-1} & u_N^{N-2}v_N & ... & u_N v_N^{N-~2}& v_N^{N-1} \\
\end{array}\right|\cr&&\cr
&=&\prod_{i<j}(u_iv_j-v_iu_j)
\end{eqnarray}
and the physical ground state wave function $\psi$ is precisely
Laughlin's wave function \cite{haldane}.

A low energy band of excited states above the $\nu=1/(2k+1)$ ground
state is constructed by promoting a composite fermion from the lowest
pseudo-Landau level to the first excited pseudo-Landau level.  On the
sphere these excited states form multiplets labeled by their total
angular momentum $l_{\rm tot}=1,...,N$.  For small systems (up to 10
electrons) Dev and Jain \cite{dev} have shown that after projection the
$l_{\rm tot} =1$ state is eliminated and the remaining band of low
energy states have large overlaps with the corresponding eigenstates
obtained by exact diagonalization.  Here I concentrate on the $l_{\rm
tot}=N$ multiplet.  The trial state for the $m_{\rm tot}=-N$ member of
this multiplet is formed by removing a composite fermion from the lowest
pseudo-Landau level at the bottom of the sphere and reintroducing it
into the first excited pseudo-Landau level at the top of the sphere.
The mean field composite fermion wave function for this state is
\begin{eqnarray}
\Phi^\prime_{CF} =
\left|
\begin{array}{ccccc}
 u_1^{N-1} & u_1^{N-2}v_1& ... & u_1 v_1^{N-2} & u_1^{N-1} {\overline v}_1 \\
\vdots & \vdots  &  &\vdots & \vdots \\
 u_N^{N-1} & u_N^{N-2}v_1 & ... & u_N v_N^{N-2} & u_N^{N-1} {\overline
v}_N \\
\end{array}\right|
\end{eqnarray}
and the corresponding physical wave function $\psi^\prime$ describes a
state with a quasielectron at the top of the sphere and a quasihole at
the bottom of the sphere.  In the $N\rightarrow\infty$ limit the
quasielectron and quasihole are infinitely separated and the excitation
energy of this state should therefore correspond to the energy gap
$\Delta$ which appears in the activated temperature dependence of
$R_{xx}$.

Before discussing the calculation of the energy gaps it is necessary to
address the problem of projecting $\psi^\prime$ onto the lowest Landau
level.  Because $\psi^\prime$ contains only a single composite fermion
in the first excited pseudo-Landau level it can be decomposed as
\begin{equation}
\psi^\prime = \psi^\prime_0 + \psi^\prime_1
\end{equation}
where $\psi^\prime_0$ is the projected state with all $N$ electrons in
the lowest Landau level and $\psi^\prime_1$ is orthogonal to
$\psi^\prime_0$ and has $N-1$ electrons in the lowest Landau level and 1
electron in the first excited Landau level.  It follows that
$T\psi^\prime_0=0$ and $T \psi^\prime_1 = E_1\psi^\prime_1$ which
implies
\begin{equation}
\psi^\prime_0 \propto (T-E_1) \psi^\prime.
\end{equation} 
The projected expectation value of a given operator $O$ can now be
expressed in terms of unprojected expectation values as
\begin{equation}
\langle O \rangle_{\rm proj.} = 
\frac{\langle (T-E_1) O (T-E_1)\rangle}
{\langle (T-E_1)^2 \rangle}.\label{projection}
\end{equation}
Equation (\ref{projection}) is ideally suited for evaluation by
variational Monte Carlo.  The inspiration for this technique is the
generalized Lanczos method of Heeb and Rice \cite{heeb} for
systematically improving a variational wave function by applying a local
operator, not necessarily the Hamiltonian, to that wave function.  Here
$\psi^\prime$ contains only a single quasielectron and so one iteration
with $T-E_1$ is sufficient to do the full projection.  However, in
principle this method can be applied to the FQHE hierarchy $(p>1)$ where
it would provide a systematic method for improving Jain's composite
fermion wave functions.

Figure \ref{chargeprofile} shows the density profile of the excited
state wave functions $\psi^\prime$ with 12 electrons for $\nu = 1/3$,
1/5, and 1/7 before and after projection onto the lowest Landau level.
In each case the charge deficit at the bottom of the sphere $(\theta =
\pi)$ is a quasihole, while the excess charge localized in a ring near
the top of the sphere $(\theta = 0)$ is a quasielectron.  While the
quasihole is unaffected by projection because it is entirely in the
lowest Landau level, the quasielectron charge becomes more localized
after projection.  This is a natural result given that the unprojected
wave function has more degrees of freedom than the projected wave
function.  Note that the effect of projection on the quasielectron grows
weaker as $\nu$ decreases reflecting the fact that the Jastrow factor in
(\ref{jain}) becomes more effective at projecting into the lowest Landau
level with increasing $k$ \cite{jain}.

Figure \ref{energygap} shows the kinetic $(T)$ and Coulomb $(V)$
contributions to $\Delta$ before and after projection onto the lowest
Landau level plotted vs.\ $1/N$ for $\nu=1/3$.  A trivial systematic
size dependence of the Coulomb contribution has been removed by
subtracting out the Coulomb energy of two point charges with fractional
charge $\pm e/(2k+1)$ at the top and bottom of the sphere,
\begin{equation}
\Delta_{\rm Coul.} = \langle V\rangle^\prime-\langle
V\rangle+\frac{(e/(2k+1))^2} {2\epsilon R}.\label{coulomb}
\end{equation}
Likewise, the kinetic energy contribution has been modified to account
for the size dependence of the energy difference between the lowest and
first excited pseudo-Landau levels,
\begin{equation}
\Delta_{\rm K.E.} = \left(\langle T\rangle^\prime-\langle T\rangle\right)/
\left(1+\frac{1}{S^*}\right).\label{kinetic}
\end{equation}
Here $\langle...\rangle$ and $\langle...\rangle^\prime$ denote
expectation values (either projected or unprojected) in $\psi$ and
$\psi^\prime$. As can be seen in Fig.~\ref{energygap} the size
dependence of the corrected energy gaps is small for large enough $N$.

It is now possible to summarize the effect of Landau level projection on
the energy gap. For $\nu=1/3$, before projection, $\Delta = \Delta_{\rm
Coul.} + \Delta_{\rm K.E.} \simeq 0.05 e^2/\epsilon l_0 +
0.16\hbar\omega_c$, consistent with the calculations of Trivedi and Jain
\cite{trivedi}. This result is clearly unphysical in the extreme
quantum limit where the only energy scale in the problem is the Coulomb
energy $\sim e^2/\epsilon l_0$.  After projection $\Delta_{\rm K.E.}= 0$
and, as shown above, the quasielectron charge has become more localized.
This in turn leads to an increase in the Coulomb energy of the
quasielectron yielding the projected energy gap $\Delta \simeq 0.1
e^2/\epsilon l_0$, consistent both with Morf and Halperin's variational
calculation \cite{morf} and with exact diagonalization studies of small
systems \cite{rezayi,fano}.  The results of similar calculations for
$\nu=1/5$ and $\nu=1/7$ are summarized in Table 1.  As for $\nu=1/3$
projection removes the kinetic energy contribution to $\Delta$ $(\propto
\hbar\omega_c$) while concentrating the quasielectron charge and
increasing the Coulomb contribution ($\propto e^2/\epsilon l_0$).  Table
1 also includes the exact diagonalization results of Fano {\it et al.}
\cite{fano} for the energy gap extrapolated to infinite system
size for $\nu=1/3$ (up to 10 electrons) and $\nu=1/5$ (up to 7
electrons).  The agreement between these energy gaps and those obtained
here demonstrates the variational validity of Jain's quasielectron wave
function for $\nu=1/3$ and 1/5. Note that because the size of the
Hilbert space needed to study systems with a given number of electrons
grows factorially with $k$, the method used here is the only efficient
way to calculate properties of projected composite fermion wave
functions, even with a single quasielectron, for $\nu = 1/7$.

To conclude, the energy gaps for the $\nu=1/3$, 1/5, and 1/7 FQHE have
been calculated by variational Monte Carlo using Jain's composite
fermion wave functions.  Results have been obtained before and after
projection onto the lowest Landau level using a novel projection
technique. Before projection a significant contribution to the energy
gap comes from the first excited Landau level. After projection this
contribution vanishes and the charge of the quasielectron becomes more
localized leading to an increased Coulomb contribution to the energy
gap.  For $\nu=1/3$ and 1/5 the projected energy gaps agree well with
previous calculations based on Laughlin's quasielectron wave function
and exact diagonalization studies of small systems. This agreement
establishes the variational validity of Jain's quasielectron wave
function and lends support to the composite fermion theory.  Clearly it
would be interesting to generalize the projection technique introduced
in this paper to study the FQHE hierarchy.

I would like to acknowledge useful discussions with N.\ Trivedi, E.\
Heeb, C.S.\ Hellberg, T.-L. Ho, and J.R.\ Schrieffer. This work was
supported by NSF grant No.\ DMR-92-22682 and by the National High
Magnetic Field Laboratory at Florida State University.

\begin{table}
\caption{
Unprojected and projected energy gaps for the FQHE with $\nu = 1/3$,
1/5, and 1/7 calculated using Jain's composite fermion wave functions.
The extrapolated exact diagonalization results of Fano {\it et al.}
\protect\cite{fano} are given for comparison.  Variational Monte Carlo
results are for 42 electrons.}
\label{table}
\begin{tabular}{ccccc}
& $\Delta_{\rm K.E.}$ & $\Delta_{\rm Coul.}$ & $\Delta_{\rm Coul.}$ & $\Delta$ \\ 
$\nu$  &  (unprojected)    & (unprojected)   & (projected) & (Ref.~\cite{fano}) \\ 
 & $(\hbar\omega_c)$ &   $(e^2/\epsilon l_0)$   &  
$(e^2/\epsilon l_0)$  & $(e^2/\epsilon l_0)$    \\
\tableline
1/3     &  0.163(2)         & 0.048(2)         & 0.106(3)  & 0.1036(2)  \\
1/5     &  0.082(2)         & 0.014(2)         & 0.025(3)  & 0.0244(3)  \\
1/7     &  0.053(2)         & 0.006(2)         & 0.011(3)  &   ---      \\
\end{tabular}
\end{table}

\begin{figure}
\caption{
Density profile of the $l_{\rm tot}=N$, $m_{\rm tot}=-N$ excited state
wave functions with $\nu=1/3$, 1/5, and 1/7 for a system with 12
electrons before and after projection onto the lowest Landau level.  For
each case the quasihole at the bottom of the sphere ($\theta = \pi$) is
unaffected by projection, while the quasielectron charge at the top of
the sphere ($\theta = 0$) becomes more localized.}
\label{chargeprofile}
\end{figure}

\begin{figure}
\caption{
Kinetic and Coulomb contributions to the energy gap vs.\ $1/N$ for
$\nu=1/3$ calculated using Jain's composite fermion wave functions
before and after projection onto the lowest Landau level.  Before
projection $\Delta_{\rm K.E.}\ne 0$ indicating a nonzero overlap with
the first excited Landau level.  After projection $\Delta_{\rm K.E.} =
0$ (not shown) and $\Delta_{\rm Coul.}$ has increased because the
quasielectron charge has become more localized.}
\label{energygap}
\end{figure}

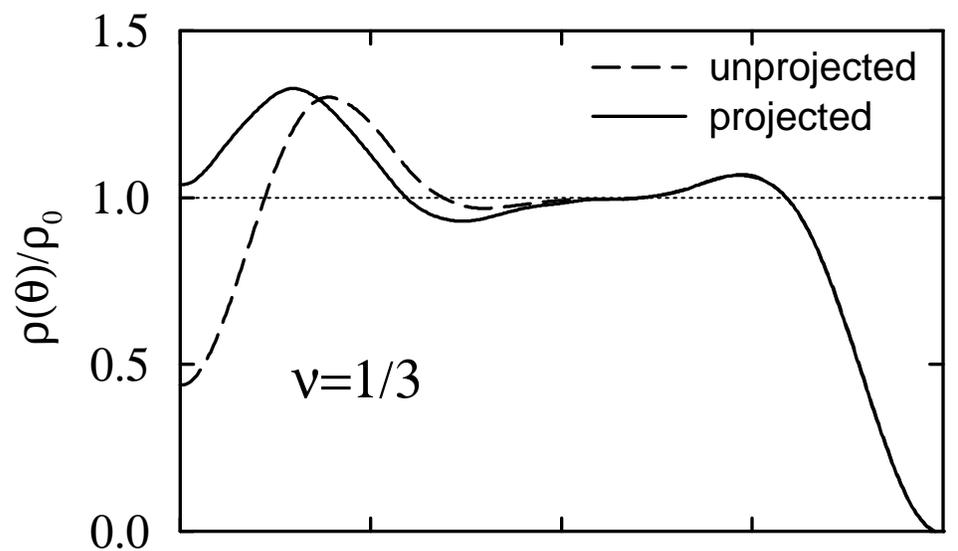
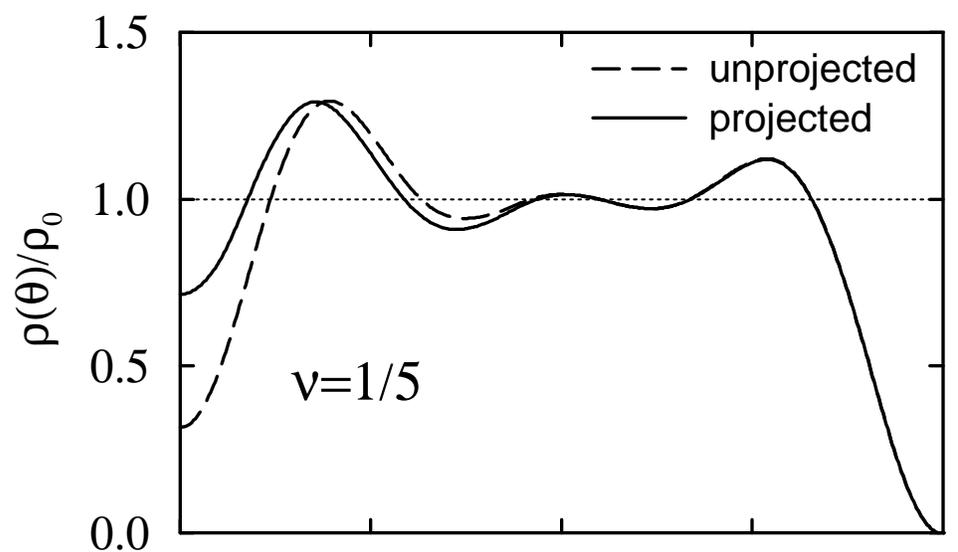
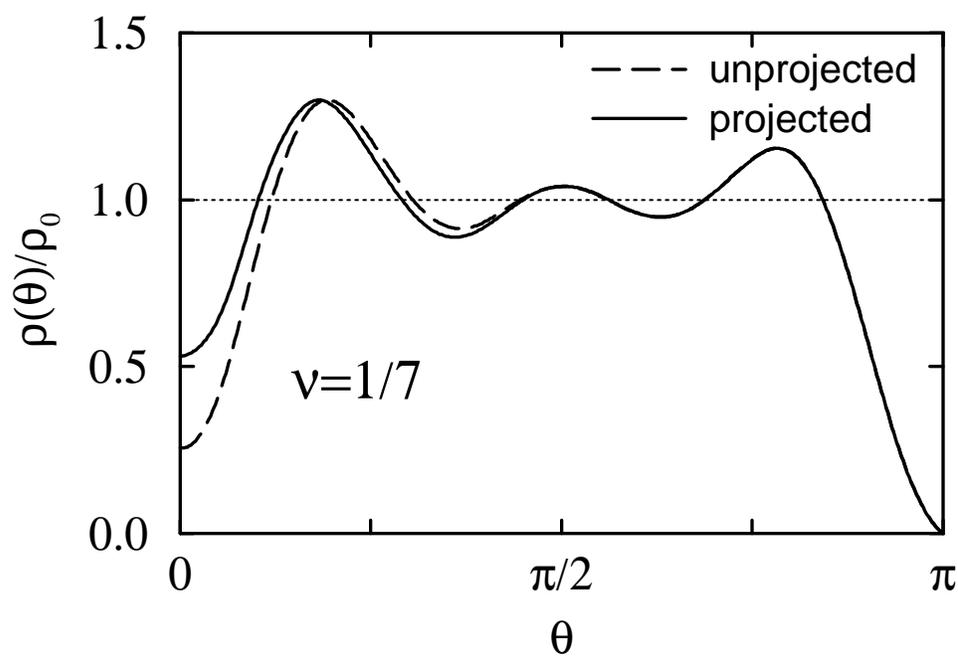
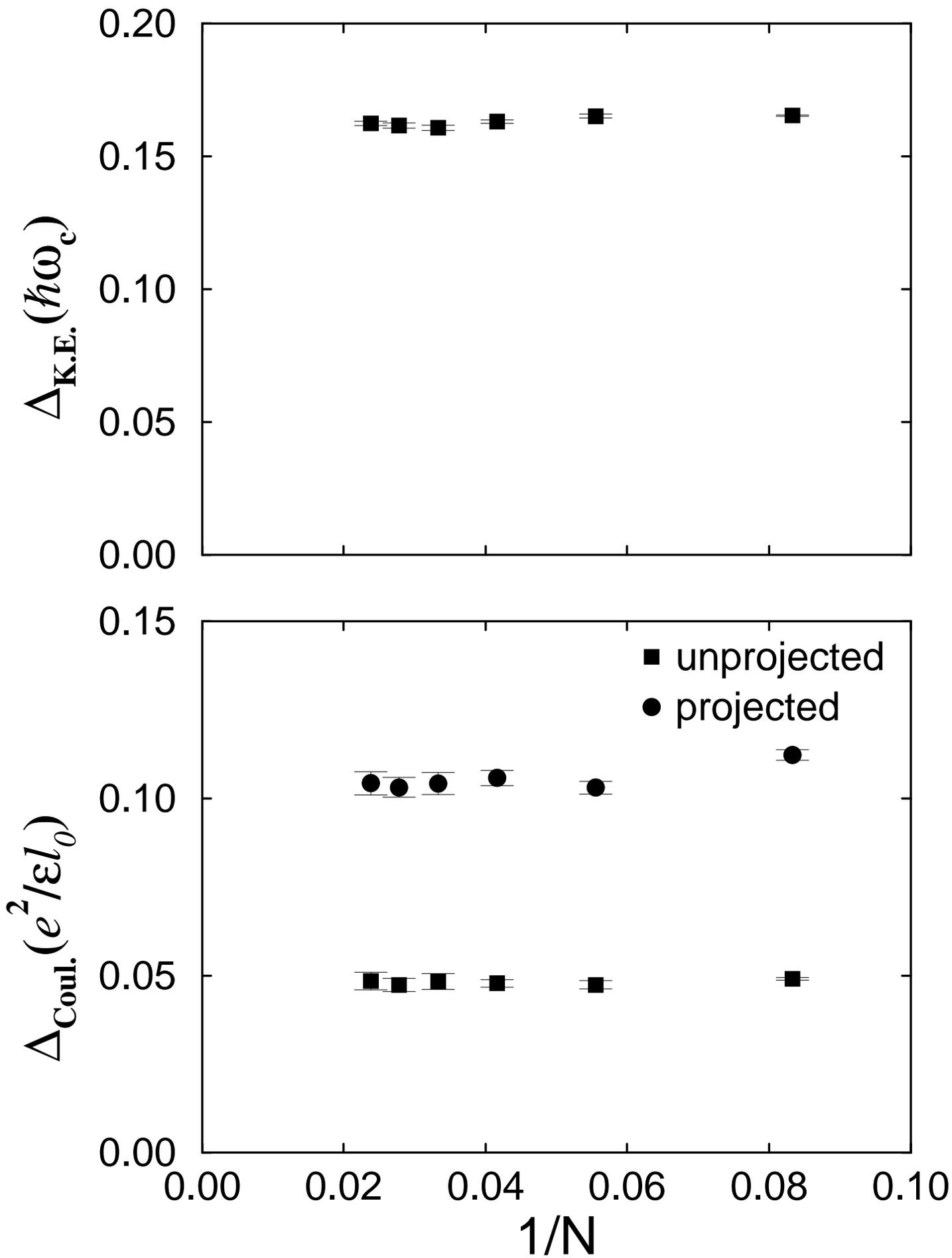
\end{document}